\documentclass[%
 aip,
 amsmath,amssymb,
 reprint,%
]{revtex4-2}

\usepackage{graphicx}
\usepackage{dcolumn}
\usepackage{bm}
\usepackage[utf8]{inputenc}
\usepackage[T1]{fontenc}
\usepackage{mathptmx}

\usepackage{etoolbox}

\makeatletter
\def\@email#1#2{%
 \endgroup
 \patchcmd{\titleblock@produce}
  {\frontmatter@RRAPformat}
  {\frontmatter@RRAPformat{\produce@RRAP{*#1\href{mailto:#2}{#2}}}\frontmatter@RRAPformat}
  {}{}
}%
\makeatother

\begin{document}

\preprint{AIP/123-QED}

\title{Bandgap widening and resonating mass reduction through wave locking}

\author{Luca~Iorio}
\author{Jacopo Maria De Ponti}
\author{Alberto Corigliano}
\author{Raffaele Ardito}%
 \email{raffaele.ardito@polimi.it.}

\affiliation{Dept. of Civil and Environmental Engineering, Politecnico di Milano, Piazza Leonardo da Vinci, 32, 20133 Milano, Italy} 

\begin{abstract}
Elastic metamaterials made from locally resonant arrays have been developed as effective ways to create band gaps for elastic or acoustic travelling waves. They work by implementing stationary states in the structure that localise and partially reflect waves. A different, simpler, way of obtaining band gaps is using phononic crystals, where the generated band gaps come from the periodic reflection and phase cancellation of travelling waves. In this work a different metamaterial structure that generates band gaps by means of coupling two contra-propagating modes is reported. This metamaterial, as it will be shown numerically and experimentally, 
generates larger band gaps with lower added mass, providing benefits for lighter structures.
\end{abstract}

\maketitle

In the last decades, the study and development of metamaterials both in the photonic and phononic world has become an interesting topic for researchers \cite{Pendry1,Pendry2,Craster1,Liu2000, Review}. The curiosity for these novel structures comes from their intrinsic wave manipulation abilities even when the wavelength interacting with the structure is much larger than the lattice size. This enables novel properties for vibration attenuation, wave guiding, wave localization and even filtering \cite{laude,Erturk2021,DePonti2021,DePonti2021Frontiers,Luca_D,Colquitt2017}. On the topic of vibration attenuation, numerous works have shown the potential and the advantages of creating what are now known as locally resonant materials \cite{Liu2000,Sun,Yu,Xiao,Erturk}. These structures generally consist of a main waveguide, that can be 1D (a beam) 2D (a plate) or 3D (a bulk solid), inside of which local resonant structures are positioned. These local structures are capable of interacting with travelling waves by means of localising the traversing energy and creating a band gap at the frequency associated with their own resonance. It has been shown also, with analytical evaluations, that the extent of the band gap generated by these local structures mainly depends on the mass ratio between the locally resonant system (generally peripheral attachments) and the mass of the main waveguide structure. Moreover, the attenuation efficiency is heavily dependent on the number of cells employed as well as the relative stiffness between the resonators and the main structure \cite{Erturk,Hussein}; the latter feature can also enable selective filtering of waves in metaframes \cite{Palle}.
The development of new metastructures that implement exotic wave manipulation capabilities is very important for the creation of novel smart structures. New manipulation effects can modify the strength, the width and even the nature of band gaps that are introduced, granting tools for developing ever more efficient systems that can repel, localise or even damp out unwanted external or internal vibrations.

In this work, a metamaterial beam that can generate a band gap employing the so called locking effect \cite{Veering,Locking,Manconi,Ruzzene_beam,JMDP_LI} is proposed and evaluated. This different physical phenomenon, if properly implemented, can give new advantages for specific structures. For example, as it will be shown, the band gap can be greatly enlarged with respect to the simple local resonance counterpart, while reducing the added lateral mass. In the proposed design the band gap analysed is generated by the coupling of two contra-propagating modes (flexural and torsional) that interact thanks to a shift in position of the lateral resonators. 
This structure can significantly cut costs with regards to adding mass to the structure, giving also the benefit of having lighter metaframes.
This phenomenon can also be looked at from a different prospective, by considering inertial amplification  \cite{zaccherini2021stress}. As a matter of fact, while for the local resonance structure the inertia is driven by translational motion due to flexure, this is enhanced in the case of locking by the rotational contribution introduced by torsion.  

\begin{figure}[t!] 
	\centering
	\includegraphics[width=0.4\textwidth]{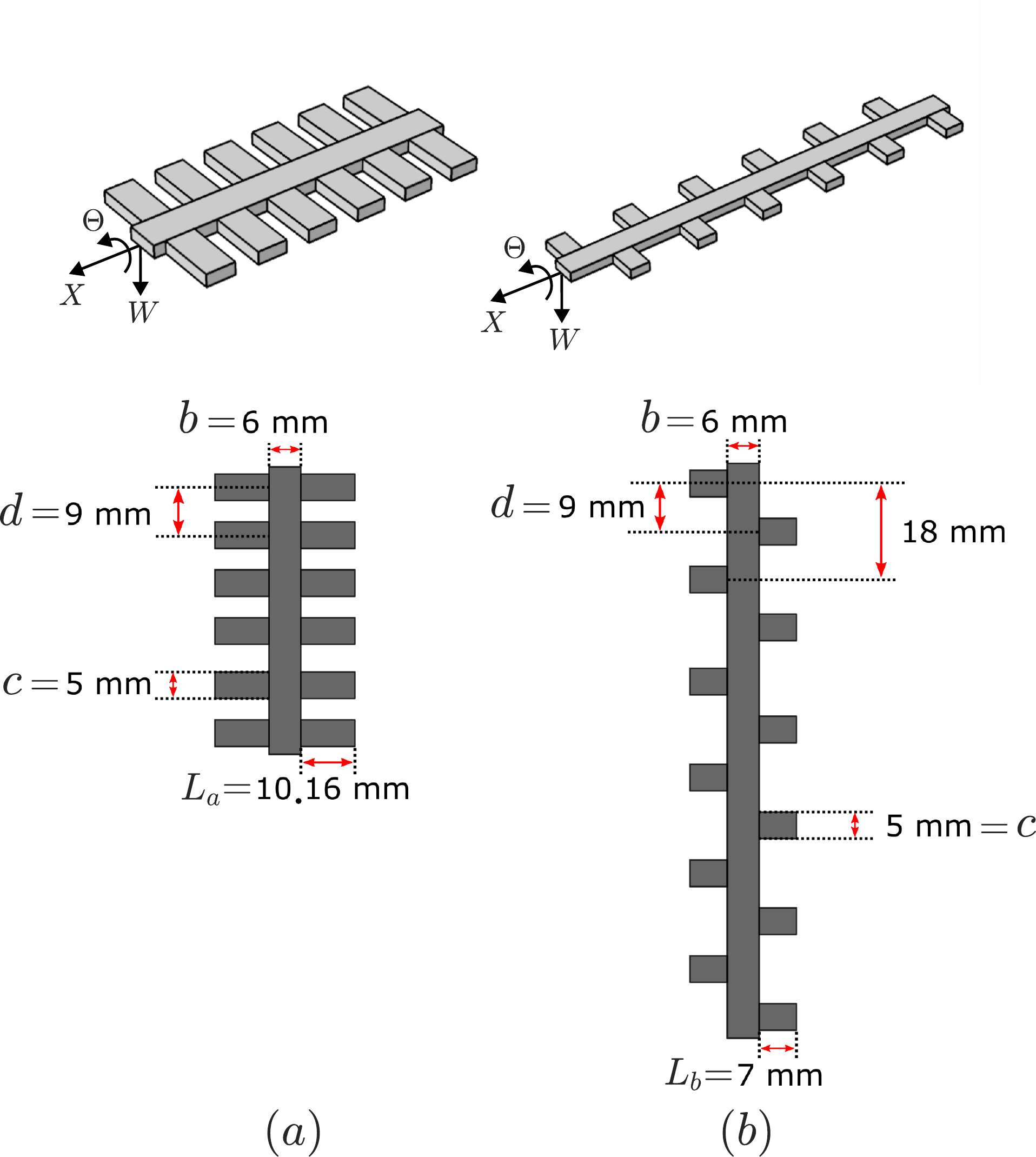}
	\caption{Geometries analysed in this work: local resonance geometry (a) and locking geometry (b), with their respective vertical elevation. The thickness of both systems is $h=2$ mm.}
\label{fig:geometria}
\end{figure}

\begin{figure*}[t!] 
	\centering
	\includegraphics[width=1\textwidth]{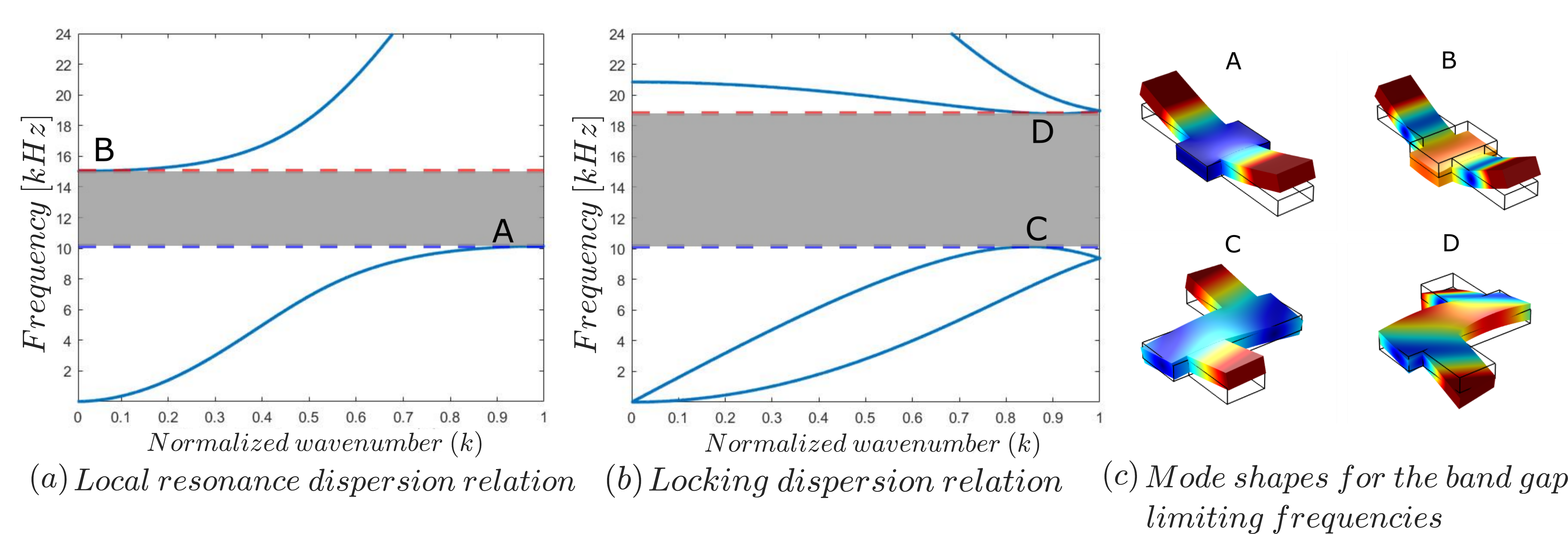}
	\caption{a) Dispersion relation for the local resonance geometry. b) Dispersion relation for the locking geometry. The opening and closing of the band gaps are reported as dashed colored lines, while in light grey the band gap zone is highlighted. c) Reports the mode shape associated to the highlighted frequencies.}
\label{fig:dispersione}
\end{figure*}

As stated above, the band gap developed for this metamaterial stems from the implementation of the locking effect in the main waveguide beam.
Locking effect is the physical phenomenon that arises from the coupling of two contra-propagating waves with different polarisation. The coupling is responsible for the generation of at least one point of zero group velocity inside the first Brillouin zone and a band gap. The band gap is obtained through the looped energy conversion between one wave mode into another with opposite propagating direction. This  determines a complete reflection of the propagating wave energy if the frequency is higher than the one associated to the zero group velocity point \cite{Manconi,JMDP_LI}. The local resonance, which is at the base of the commonly engineered passive metamaterials, on the other hand, arises from the coupling between a propagating mode and a stationary mode.
In this paper, we compare two similar devices that show the aforementioned physical phenomena. In both cases, the waveguide is represented by an Euler-Bernoulli beam, on which resonators are attached in the form of cantilever beams. To achieve classical local resonance, the resonators are located in a symmetric fashion with respect to the beam's axis, see Fig.\ref{fig:geometria}(a); conversely, wave locking can be achieved by introducing alternate resonators on a single side of each cell, see Fig.\ref{fig:geometria}(b). The analytical modelling for the evaluation and prediction of the band gaps generated by the two different physical phenomena has already been discussed in previous works \cite{JMDP_LI,Skelton} and the equations are used to match the opening of the two band gaps. More specifically, we consider that, for the sake of simplicity, the motion of the resonator is dominated by the first eigenmode of the cantilever. Consequently, the dynamic equilibrium of each resonator is established in terms of the equivalent bending stiffness $k$ and the participating mass $m$. The reaction force $F_n$ and the reaction moment $M_n$ at the attachment point of each resonator are given by:
\begin{equation}
\begin{split}    
    &F_n=k \left[\psi_n-\left(w_n+(-1)^n \theta_n \frac{b+\hat{L}}{2} \right)\right]  \\[4pt]
    &M_n=F_n \frac{\hat{L}}{2},
\end{split}
\label{eq:01}
\end{equation}
where: $\psi_n=\psi_n(t)$ is the degree of freedom that describes the motion of the $n$-th resonator; $w_n=w(nd,t)$ is the transverse displacement of the waveguide in correspondence of the $n$-th resonator; $\theta_n=\theta(nd,t)$ is the torsional rotation of the waveguide in correspondence of the $n$-th resonator. The equation of motion for each resonator is written as:
\begin{equation}
\frac{\partial^2 \psi_n}{\partial t^2} + \omega^2_0  \left[\psi_n-\left(w_n+(-1)^n \theta_n \frac{b+\hat{L}}{2} \right)\right] =0.
\label{eq:01b}
\end{equation}
The eigenfrequency of the resonator is denoted by $\omega_0=\sqrt{k/m}$.

The equation of motion of the waveguide is written in different form for the two device. For the case of classic local resonance, in view of the symmetric distribution of resonators, the torsional rotation is null, whereas the equilibrium in the transverse direction contains the contribution of the two resonators in each cell:
\begin{equation}
\begin{split}
    &EI\frac{\partial^4 w}{\partial x^4}+ \rho A\frac{\partial^2 w}{\partial t^2}=\sum_{n=-\infty}^{n=+\infty}2F_n \delta (x-nd)  \\[4pt]
    &\theta=0.
\end{split}
\label{eq:02}
\end{equation}
In the above equation, $E$ is the Young's modulus and $\rho$ is the mass density of the material of the waveguide; $A$ and $I$ represent the area and the moment of inertia of its cross-section, respectively; $\delta$ is the Dirac's delta.

In the case of wave locking, the asymmetric configuration entails the coupling with torsional behavior, so that the equations of motion read:
\begin{equation}
\begin{split}
    &EI\frac{\partial^4 w}{\partial x^4}+ \rho A\frac{\partial^2 w}{\partial t^2}=\sum_{n=-\infty}^{n=+\infty}F_n \delta (x-nd)  \\[4pt]
    &GJ\frac{\partial^2 \theta}{\partial x^2}- \rho I_p\frac{\partial^2 \theta}{\partial t^2}=-\sum_{n=-\infty}^{n=+\infty} (-1)^n F_n \frac{b+\hat{L}}{2}\delta (x-nd) .
\end{split}
\label{eq:03}
\end{equation}
where: $G$ is the tangential modulus of the material; $I_p$ and $J$ are the polar moment of inertia and the (primary) torsional stiffness of the cross-section, respectively. 

\begin{figure*}[t!] 
	\centering
	\includegraphics[width=1\textwidth]{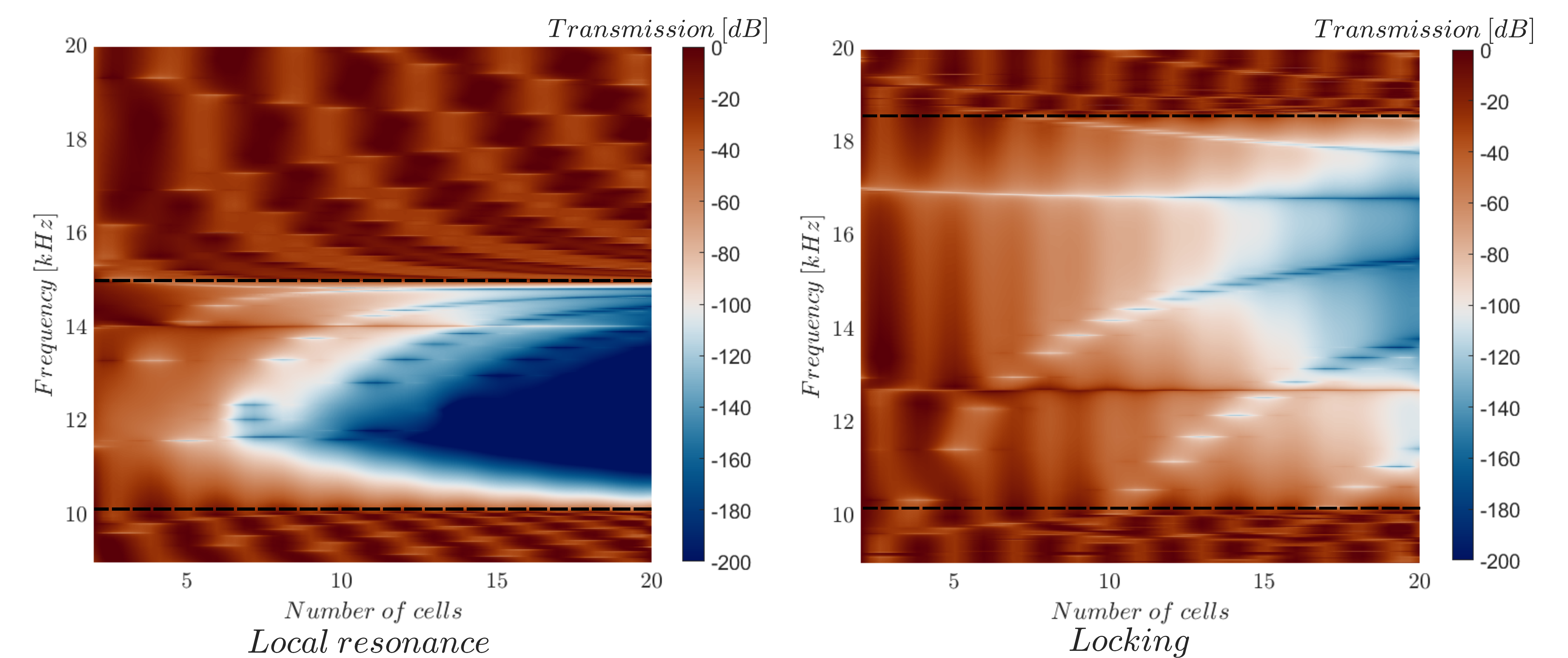}
	\caption{Evaluation of the attenuation granted by the two metamaterials in the band gap region, varying the number of cells considered. Dashed black horizontal lines report the theoretical opening and closing of the band gaps}
\label{fig:band_gap}
\end{figure*}

The systems of equations (\ref{eq:01}), (\ref{eq:01b}), (\ref{eq:02}) and (\ref{eq:01}), (\ref{eq:01b}), (\ref{eq:03}) can be solved by adopting the plane-wave expansion method (PWEM), as explained in a previous paper\cite{JMDP_LI}. The analytical solution is useful in order to design the devices with the same opening frequency of band gap. As shown in Fig.\ref{fig:geometria}, we assume the same cross-section for the waveguide and the same dimension $d$ of the unit cell. The cross-section of the lateral resonators is also the same, but their length is different. We assume a predefined value $L_b$ for the system that shows wave locking and, by examining the analytical dispersion curve, we obtain the value $L_a$ in order to match the band gap opening frequency. As a result, we obtain that for wave locking the resonators are 31.10\% shorter than the classical resonance case.

The numerical analyses now reported are all conducted using the commercial software COMSOL Multyphysics, with the 3-dimensional modelling of the devices shown in Fig.\ref{fig:geometria}. The material used for both configurations is aluminum ($E$ = 70 GPa, $\rho = 2710 $ kg/m$^{3}$, $\nu = 0.33$). 

Fig.\ref{fig:dispersione} reports the dispersion relation for the two geometries. In the locking case there is a clear coupling between the first flexural mode and the first torsional mode and this in turns generates the band gap. The coupling is granted by the shift in position of the resonators breaking the symmetry of the cell. The locking effect is more complicated to engineer with respect to the local resonance or the Bragg scattering of a phononic crystal, because of the need to tailor the cell geometry (the resonator, the length of the cell and so on) so that in the reciprocal space there is a liftable accidental degeneracy of two contra propagating modes. This is particularly difficult to obtain for low frequencies.
The dispersion shows that the opening frequency for the band gap of the two geometries is the same (10156 Hz) while the closing frequency is very different: for the local resonance geometry the closing frequency is 15012 Hz while for the locking geometry it closes at 18788 Hz. 
The gap-mid gap ratio is then evaluated to obtain a non-dimensional parameter that avoids
frequency dependence. The formula is:
\begin{equation}\label{gmgr}
BG = \frac{2(f_{top}-f_{bot})}{f_{top}+f_{bot}} \cdot 100
\end{equation}
where $f_{bot}$ and $f_{top}$ are the frequencies that delimit the band gap.
From equation (\ref{gmgr}) it is obtained that the local resonance case has a gap-mid gap ratio of 38.56\%, while for the locking it is 59.65\%. Given that the overall added mass per cell is reduced by 31.10\% this is an improvement in the effectiveness at creating bigger band gaps, reducing the added mass to the structure.
These results may be explained by considering that the actual resonant masses in the two systems are different. The local resonance has just the lateral resonators as participants of the resonant motion, effectively being the only resonant masses of the modes. The locking case, on the other hand, employs not only the mass of the lateral resonators, but also the entire structure in the motion of the evaluated modes.

\begin{figure}[h!] 
	\centering
	\includegraphics[width=0.4\textwidth]{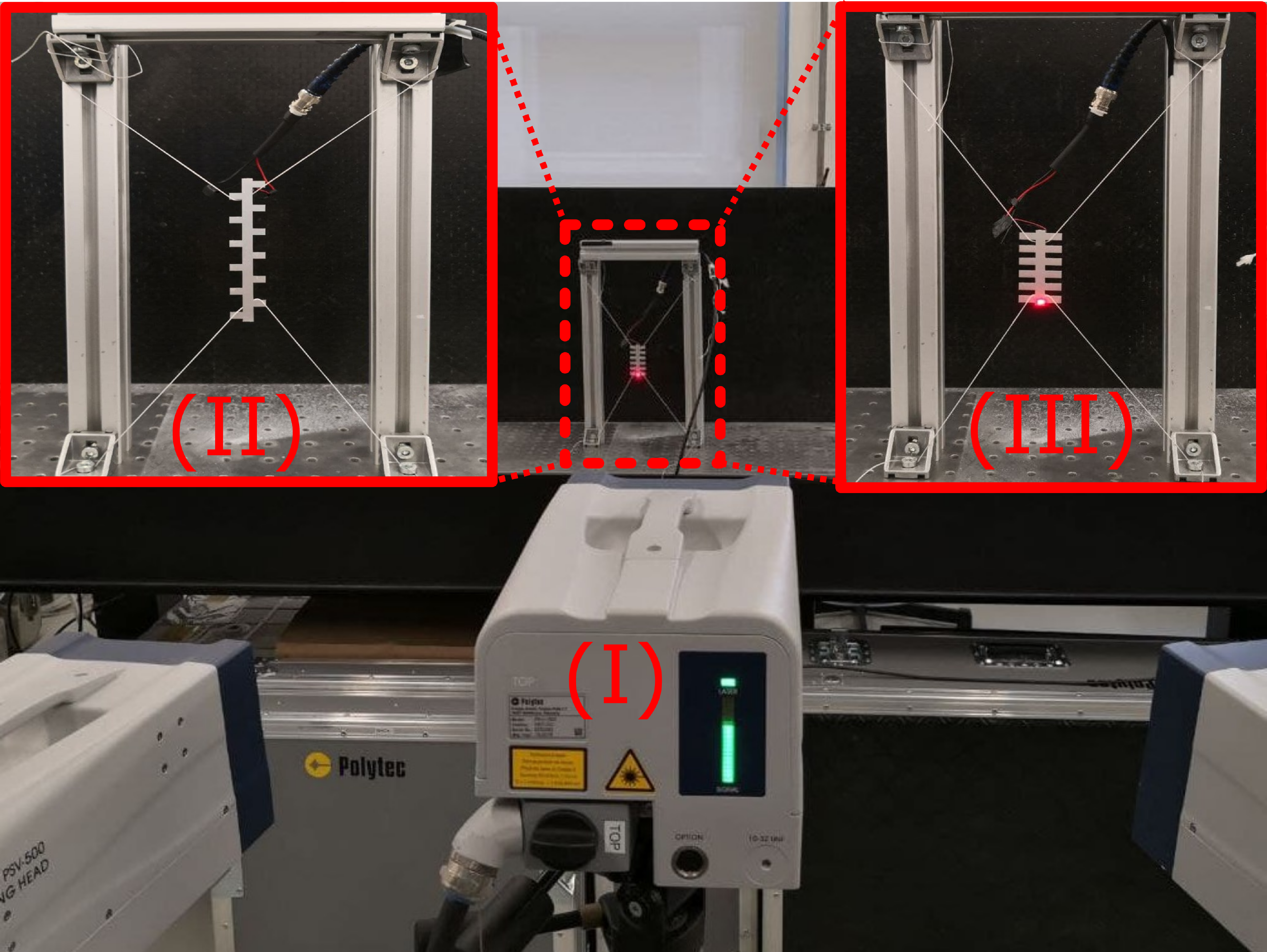}
	\caption{Picture of the experimental setup: (I) marks one of the three heads of the vibrometer, (II) is the locking specimen, while (III) is the local resonance specimen.}
\label{fig:experiment}
\end{figure}

\begin{figure*}[t!] 
	\centering
	\includegraphics[width=1\textwidth]{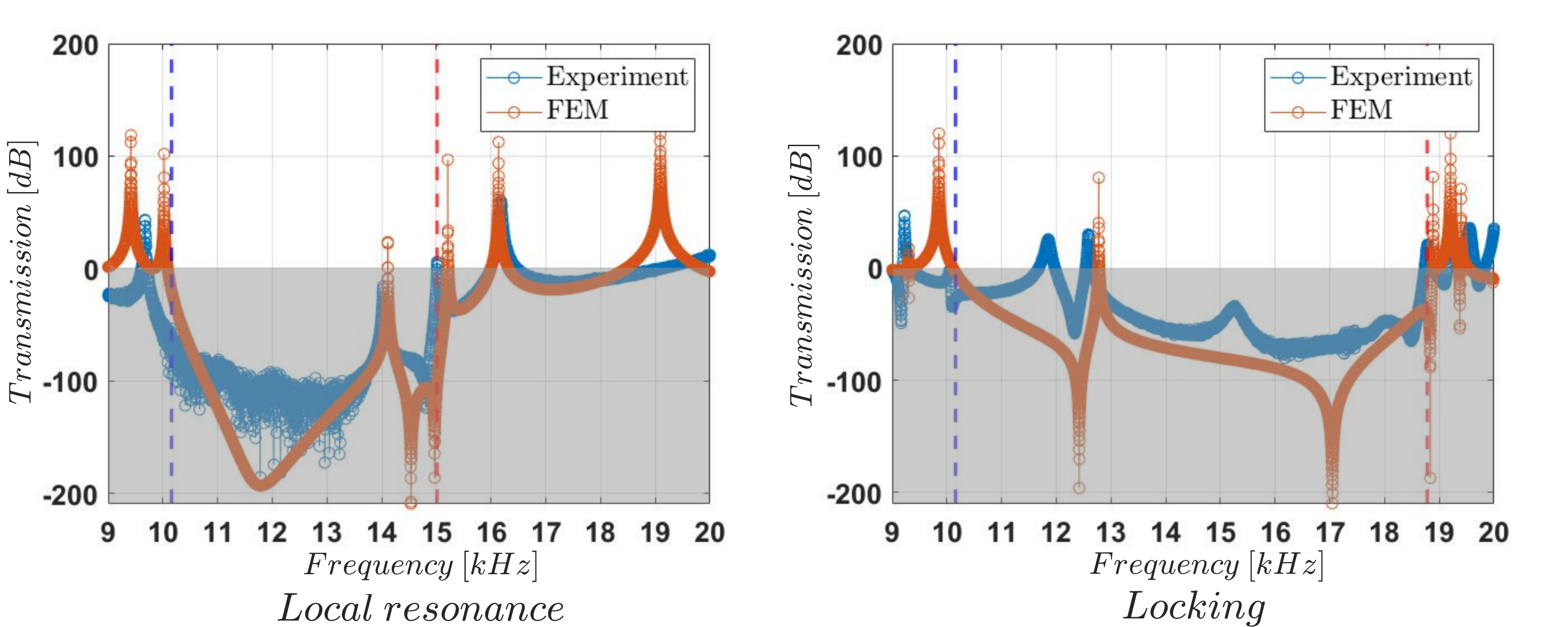}
	\caption{Evaluation of the transmission in dB for the two structures subjected to the same sweep in frequency. We reported both the experimental outcomes solutions and the numerical results obtained with the finite element methods.}
\label{fig:db_exp}
\end{figure*}

Fig.\ref{fig:band_gap} now reports the analyses conducted to evaluate the effectiveness of the band gap. The attenuation capability given by the metamaterial at different frequencies is evaluated by means of a frequency sweep in the frequency domain analysis. Combining it with a second sweep of the number of cells provides us a clear indication on both where the band gap is and how many cells are needed to obtain an effective attenuation. We conclude that the attenuation capability of the locally resonant material is significantly higher with respect to the locking case for the same number of cells. This means that even though the band gap may be bigger, the attenuation is less efficient. As stated in previous research\cite{Erturk}, for a small number of cells the band gap is larger with respect to the one defined by the dispersion relation (even though the attenuation is almost negligible), then, increasing the number of cells, the band gap shrinks and its opening and closing frequencies become defined. This is also true for the locking configuration, but it seems that the number of cells required for the stability of the band gap is lower, given that the boundaries are well defined already at eight cells. 

To test the results obtained though finite element simulations, experiments with real specimens were also performed. The specimens are fabricated by means of laser cutting an aluminum plate of the desired thickness. The number of elementary cells that were sufficient for the desired properties is six. This was done analysing the preliminary numerical results. The experimental setup is as follows: the specimens are hanged from vertical supports to have free ended boundary conditions. A piezoelectric patch PZT-5H ($ E_p $ = 61 GPa , $\nu_p = 0.31 $, $\rho_p = 7800$ kg/m$^{3}$ with dielectric constant $ \epsilon^{T}_{33}/\epsilon_0 = 3500 $ and piezoelectric coefficient $\epsilon_{31} = -9.2$ C/m$^{2}$) is attached to one end and it is used to supply the flexural excitation. The attenuation is then evaluated by extracting the out-of-plane velocity of the points at the center line of the metabeams thanks to a Polytec 3D Scanner Laser Doppler Vibrometer (SLDV). Fig. \ref{fig:experiment} shows the setup of the laboratory and the specimens. The results are reported in Fig.\ref{fig:db_exp} for both metastructures correlated with their respective finite element simulations. It is clear that the band gaps predicted by the numerical simulations are also present in the specimens produced. There are however some discrepancies between the FEM model and the actual specimens, mainly due to the presence of resonant modes. This can be attributed to small imperfections in the specimens, non perfect boundary conditions and finally the presence of the piezoelectric patch.

In conclusion, a design concept for a vibration isolation meta-beam has been reported. With it, the analyses have shown that the subtraction of mass and the breaking of the symmetry in the position of the resonators grants the creation of larger, but slightly less efficient band gaps. This works even for structures that employ a limited number of cells. The downside of this lower attenuation capability is given by the significantly lower added mass. Furthermore the cell geometry is more limiting with respect to the local resonance case given the need of bigger cells and the more complicated engineering of the modes that have to be bent in the right way in order to obtain the locking effect and consequentially the band gap. Further studies of these locking designs could lead to new generations of metalattices and metaframes for further advancements of vibration isolation frames.

\begin{acknowledgments}
The support of the H2020 FET-proactive project MetaVEH under grant agreement No. 952039 is acknowledged.
\end{acknowledgments}

\section*{Data Availability Statement}
The data that support the findings of
this study are available from the
corresponding author upon reasonable
request.


\nocite{*}  
\bibliography{Main}

\end{document}